\documentclass[showpacs,floatfix,a4paper,prl,twocolumn]{revtex4}
\usepackage{amsfonts}
\usepackage{amsmath}
\usepackage{graphicx}
\usepackage{latexsym,epsfig,bm,times,psfrag,subfigure}
\usepackage{bm,times}
\usepackage{dcolumn}

\begin{document}

\title{Relaxation in driven integer quantum Hall edge states}
\author{D.\ L.\ Kovrizhin$^{1}$ and J.\ T.\ Chalker$^{2}$}
\affiliation{$^1$Max Planck Institute for the Physics of Complex Systems, N\"{o}thnitzer
str. 38, Dresden, D-01187, Germany}
\affiliation{$^2$Theoretical Physics, Oxford University, 1, Keble Road, Oxford, OX1 3NP,
United Kingdom}
\date{\today}
\pacs{71.10.Pm, 73.23.-b, 73.43.-f, 42.25.Hz}

\begin{abstract}
A highly non-thermal electron distribution is generated when quantum Hall edge states
originating from sources at different potentials meet at a quantum point contact. 
The relaxation of this distribution to a stationary form as a function of distance downstream from the contact
has been observed in recent experiments [C. Altimiras {\it et al.} Phys. Rev. Lett.~\textbf{105},~056803~(2010)].
Here we present an exact treatment of a minimal model for the system at 
filling factor $\nu{=}2$, with results that account well for the observations.
\end{abstract}

\maketitle


{\it Introduction.}
The importance of understanding non-equilibrium dynamics and relaxation in
many-body quantum systems has been recognised since the early years of
quantum mechanics \cite{vonNeuman,review}. 
Settings in which such problems are of high current interest include, among
others, cold atomic gases \cite{review} and nanoscale electronic devices.
\cite{pierre2010,pierre2009,altimiras2010,heiblum2,granger2009,diffusive,meir,konik,andrei,kouwenhoven,fqhe-iv,chang}
As a particular example, recent experiments \cite{pierre2010} on quantum Hall (QH) edge states driven out of equilibrium at a quantum
point contact (QPC) provide very detailed information on the approach to a steady state in an electron
system that appears to be well-isolated from other degrees of freedom. In this paper we describe the 
exact solution of a simple model for these experiments and compare our results with the measurements. 

In outline, the experiments we are concerned with \cite{pierre2010} involve
two sets of integer QH edge states, which meet at a QPC. 
When a bias voltage is applied to the QPC, tunneling between the edge
states generates a non-equilibrium electron distribution. The form of this
distribution in energy and its evolution as a function of distance
downstream from the QPC are probed by monitoring the tunneling current from
a point on the edge, through a quantum dot that has an isolated level of
controllable energy. Close to the QPC, the measured distribution has two
steps, reflecting the different energies of Fermi steps in each of the
incident edges. With increasing distance from the QPC, the distribution
relaxes to a single, broad step. The theoretical challenge presented by these
observations is to understand and model this relaxation process.

The relationship between these edge state experiments and other recent work
on many-body quantum dynamics far from equilibrium has several aspects worth
emphasising. First, in the context of QH edge states, these are the most
recent of a series of striking observations of non-equilibrium effects in
interferometers \cite{heiblum2} and in thermal transport \cite{granger2009}.
They are also the equivalent for a ballistic system of earlier studies \cite%
{diffusive} of local distributions in diffusive wires. Second, the
measurements stand apart from earlier theory \cite{fqhe-iv} and experiment 
\cite{chang} on non-equilibrium transport between fractional QH edge states,
because they probe local distribution functions, rather than the global
non-linear current-voltage characteristic. Third, and more broadly, the
system studied is different in important ways from quantum impurity problems 
\cite{meir,konik,andrei,kouwenhoven}, as there is no impurity degree of
freedom and interactions are not confined to an impurity site but instead
operate in the ingoing and outgoing channels. Fourth, there is an analogy
between edge state relaxation and cold atom experiments in the time domain 
\cite{review}, since distance from the QPC translates roughly as time, using
the edge state velocity as a conversion factor. In that
sense the experiment we consider, probing relaxation as electrons propagate, 
is equivalent to a quantum quench \cite{calabrese}, in which time evolution
is studied following a sudden change in the Hamiltonian. An important
question in this context is whether a system thermalises
at long times. Since integrability is an obstacle to thermalisation, it is
noteworthy that the conventional model \cite{Wen} of a QH edge state as a
chiral Luttinger liquid is an integrable one.


Any attempt to model theoretically the experiments of Ref.~\onlinecite{pierre2010}
starting from a chiral Luttinger liquid description faces an obvious
difficulty, since interactions are most naturally described in terms of
collective modes using bosonization, but tunnelling at the QPC is simple only
in terms of fermionic variables. In pioneering work, two alternative
approaches have been developed: one based on a Boltzmann-like equation for
the electron distribution \cite{buttiker2010}; and the other using a
phenomenological model for the plasmon distribution generated at the QPC 
\cite{degiovanni2010}.  From \cite{degiovanni2010}, and from subsequent discussion of a quantum
quench in an isolated QH edge (with an initial state chosen to emulate the
effects of a QPC) by the present authors \cite{quench2009}, a physical
picture has emerged, in which relaxation is seen as a consequence of 
plasmon dispersion or the presence at $\nu=2$ of two plasmon modes with distinct velocities.
In summary, an electron that tunnels at the QPC can be viewed as a superposition of
plasmons. Dispersion or multiple plasmon velocities cause such a wavepacket
to broaden as it propagates. The lengthscale for relaxation of the electron distribution
downstream from the QPC is the distance at which the width of this wavepacket
is comparable to the characteristic separation between tunnelling electrons.
The argument identifies relevant scales but does not generate a
prediction for the electron distribution and its dependence on distance from
the QPC. One of our main aims here is to calculate this central quantity. 

An overview of our treatment is as follows: we make key use of the fact
that, since experiments are at Landau level filling factor $\nu {=}2$, each
QH edge carries two co-propagating channels. Interactions mix excitations in
different channels, and in the minimal model eigenmodes are charactised
using two velocities. Our approach combines refermionisation \cite%
{Fabrizio,zarand} with non-equilibrium bosonization methods \cite%
{cazalilla,gutman,LKnoneq,mzprb,LS2011}: by first bosonizing, then
recombining bosons to form a new set of fermions, we transform the
Hamiltonian for the interacting system into one for free particles. Under
this transformation, the measured distribution (or more accurately, the
spectral function probed by the tunnelling conductance) is expressed as a free-fermion
determinant. This determinant has a form similar to that appearing in the
the theory \cite{klich,abanin-levitov} of full counting statistics [FCS],
and related quantities appear in a variety of other non-equilibrium problems 
\cite{mirlin}. Its numerical evaluation can be done accurately and
efficiently, yielding the results we present below. In addition, simple
analytical expressions can be found for some quantities, and we give one for
the total energy in each channel as a function of distance from the QPC, thus providing a Hamiltonian
derivation of a result obtained previously in the phenomenological treatment
of Ref.~\onlinecite{degiovanni2010}.

The work we set out here is related in a variety of ways to other studies
of devices built from integer QH edge states. In particular, experiments that
revealed striking non-equilibrium effects in Mach Zehnder interferometers \cite{heiblum2}
have stimulated extensive theoretical research, \cite{chalker,neder2007,cheianov,sukh2008,neder,sim,mzprb,Mirlin,neder2012}
including calculations \cite{sukh2008} of interferometer dephasing based on the same bosonized
model \cite{Wen} for edge states at $\nu=2$ with contact interactions that we adopt in the following. 
In the context of edge state relaxation a parallel development to the present paper,
described in Ref.~\onlinecite{LS2011}, is based on an approximation 
(involving a factorisation of bosonic correlators) that is expected to be accurate 
sufficiently far from the QPC. Our solution recovers such a factorisation, but only at distances large 
compared to the relaxation length for the model. 

\begin{figure}[t]
\epsfig{file=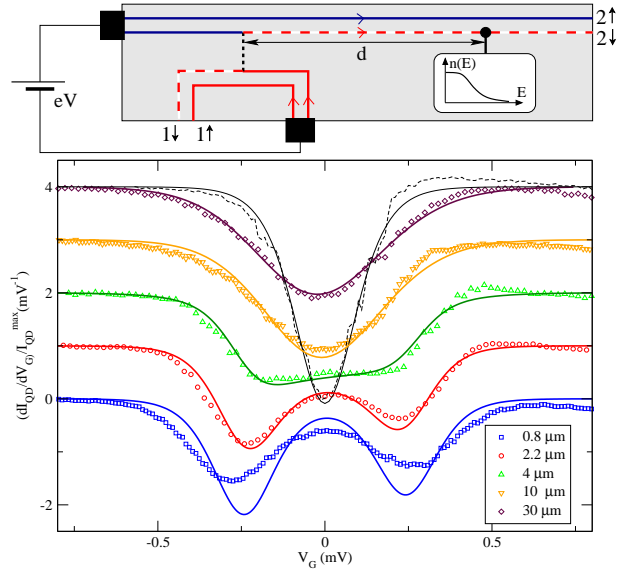,width=8cm}
\caption{(Color online) Top: sketch of the experimental geometry in which
the tunnelling conductance is measured in the \textit{same} channel as that coupled by the
QPC. Bottom: differential conductance calculated (lines) and measured \protect\cite%
{pierre2010} (symbols) in this geometry at indicated distances from QPC.
Fits use $T=44$ $\mathrm{mK,}$ $V=30.3\ \mathrm{\protect\mu V}$ ($\protect%
\beta eV=8)$, and $p=0.545.$ Black lines: calculations (full) and data
(dashed) for $V{=}0$.\label{fig1}}
\end{figure}

\begin{figure}[tbp]
\epsfig{file=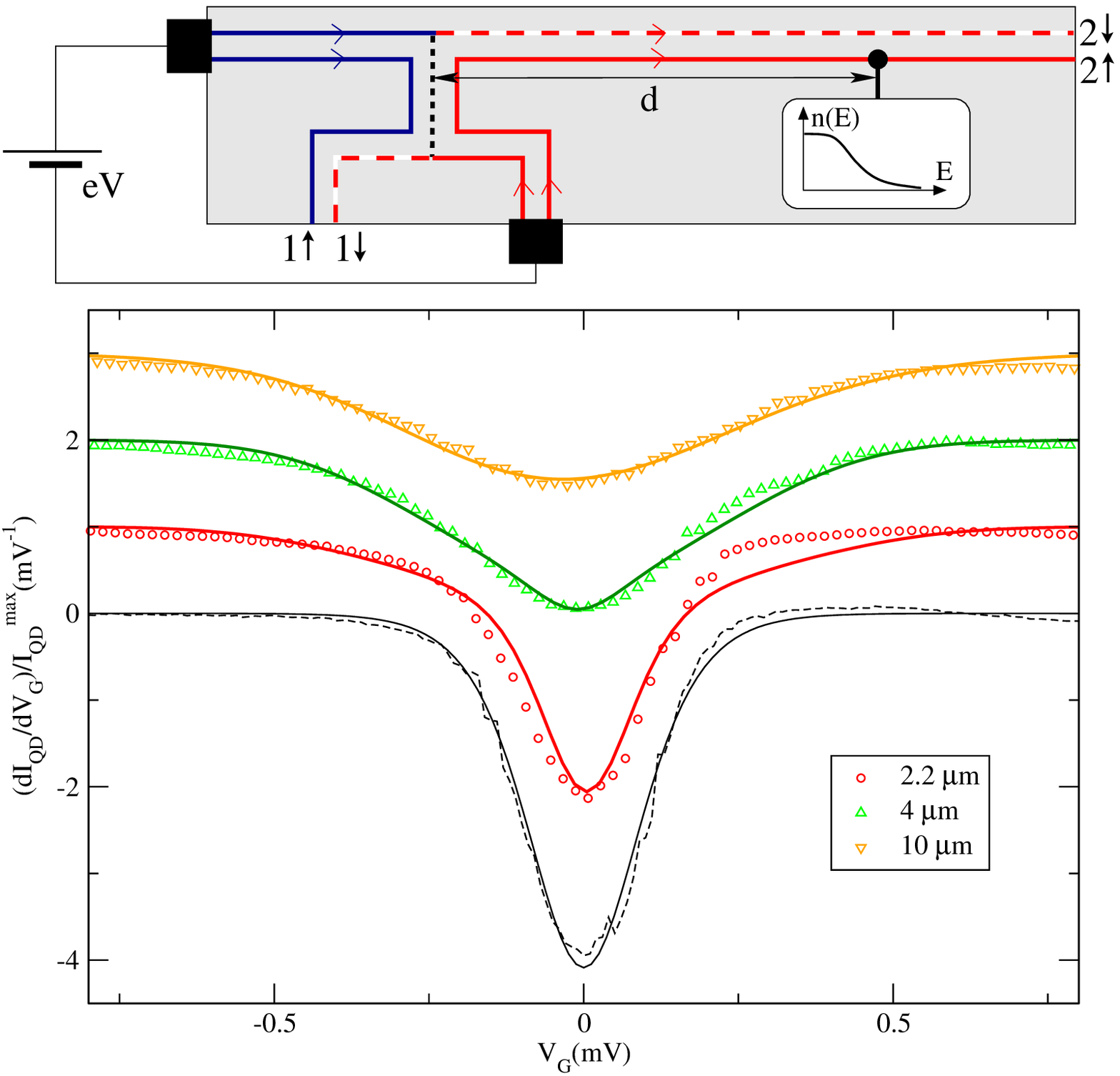,width=8cm}
\caption{(Color online) Top: sketch of the experimental geometry in which
the tunneling conductance is measured in the \textit{opposite} channel to that coupled by the
QPC. Bottom: differential conductance calculated (lines) and measured \protect\cite%
{pierre2010} (symbols) in this geometry at indicated distances from the QPC.
Fits use $T=44$ $\mathrm{mK,}$ $V=45.5\ \mathrm{\protect\mu V}$, and $p=0.545
$. Black lines: calculations (full) and data (dashed) for $V=0$.\label{fig2}}
\end{figure}

{\it Model.}
The experimental system is illustrated in the upper panels of Figs.~\ref{fig1} and \ref{fig2}:
two alternative geometries arise, according to whether the tunnelling conductance is measured
in the \textit{same} or the \textit{opposite} channel to that coupled by the
QPC.

We take a Hamiltonian with kinetic, interaction and tunneling terms, in the
form $\hat{H}=\hat{H}_{\mathrm{kin}}+\hat{H}_{\mathrm{int}}+\hat{H}_{\mathrm{%
tun}}$. Using the labels $\eta =1,2$ to distinguish edges according to their
source, and $s=\uparrow ,\downarrow $ to differentiate between the two
channels on a given edge, the fermion creation operator at point $x$ for
channel $\eta ,s$ is $\hat{\psi}_{\eta s}^{\dagger }(x)$. It obeys the
standard anticommutation relation 
$\{\hat{\psi}_{\eta s}^{\dagger }(x)\hat{\psi}_{\eta ^{\prime }s^{\prime
}}(x^{\prime })\}=\delta _{ss^{\prime }}\delta _{\eta \eta ^{\prime }}\delta
(x-x^{\prime })$.
The density operator is 
$\hat{\rho}_{\eta s}(x)=\hat{\psi}_{\eta s}^{\dagger }(x)\hat{\psi}_{\eta
s}(x)$.
Taking all four channels $\eta ,s$ to have the same bare velocity $v$, and
assuming a contact interaction of strength $g$ between electrons in
different channels on the same edge, we have \cite{Wen} 
\begin{equation}
\hat{H}_{\mathrm{kin}}=-i\hbar v\sum_{\eta ,s}\int \hat{\psi}_{\eta
s}^{\dagger }(x)\partial _{x}\hat{\psi}_{\eta s}(x)\ \mathrm{d}x
\end{equation}%
and 
\begin{equation}
\hat{H}_{\mathrm{int}}=2\pi \hbar g\sum_{\eta }\int \hat{\rho}_{\eta
\uparrow }(x)\hat{\rho}_{\eta \downarrow }(x)\ \mathrm{d}x\,.
\end{equation}%
Note that short-range intrachannel interactions can simply be absorbed into
the value of $v$. 
Tunneling with amplitude $t_{\mathrm{QPC}}$ at the QPC between
channels $1{\downarrow }$ and $2{\downarrow }$ is described by 
\begin{equation}
\hat{H}_{\mathrm{tun}}=t_{\mathrm{QPC}}\hat{\psi}_{1\downarrow }^{\dagger
}(0)\hat{\psi}_{2\downarrow }(0)+\mathrm{h.c.}\,.
\end{equation}%
A bias voltage $V$ generates a chemical potential difference $eV$ between
incident electrons on edge 1 and those on edge 2.

The observable of interest is the tunnelling conductance as a function of energy $E$ and
distance $d>0$ from the QPC, in channel $2,s$. This is the Fourier
transform of the correlator 
\begin{equation}\label{defG}
G_{s}(d,\tau )=\langle e^{i\hat{H}\tau/\hbar} \hat{\psi}_{2s}^{\dagger }(d)e^{-i\hat{H}\tau/\hbar}\hat{\psi}%
_{2s}(d)\rangle ,
\end{equation}%
where the average is taken in the non-equilibrium steady state, with
$s{=}{\downarrow }$ in the geometry of Fig.~\ref{fig1} and $s{=}{\uparrow }$ in
that of Fig.~\ref{fig2}. The tunnelling conductance is
determined by a measurement of the current $I_{\mathrm{QD}}(E)$ through a
quantum dot with a single level at energy $E$ weakly coupled to the channel.
With tunnelling amplitude $t_{\mathrm{D}}$ to the dot, this is \cite%
{quench2009} 
\begin{equation}
I_{QD}(E)=\frac{e|t_{\mathrm{D}}|^{2}}{\hbar ^{2}}\int G_{s }(d,\tau
)e^{-iE\tau /\hbar }\ d\tau .  \label{IQD}
\end{equation}%

{\it Results.}
We show in \cite{supplementary} that the Hamiltonian $\hat{H}$ can be brought into a free
particle form by introducing transformed fermion operators $\hat{\Psi}%
_{\alpha }(x)$ with index $\alpha =A_{\pm },S_{\pm }$
and eigenmode velocities $v_{\alpha }\equiv v_{\pm }=v\pm g$.
Using these new
operators 
\begin{eqnarray}
\hat{H} &=&-i\hbar \sum_{\alpha }v_{\alpha }\int \hat{\Psi}_{\alpha
}^{\dagger }\partial _{x}\hat{\Psi}_{\alpha }dx  \notag  \label{H} \\
&&-[t_{\mathrm{QPC}}\ \hat{\Psi}_{A{+}}^{\dagger }(0)\hat{\Psi}_{A{-}}(0)+%
\mathrm{h.c.}]\,.
\end{eqnarray}
Moreover the tunnelling conductance is obtained from
\begin{equation}
G_s(d,\tau )\propto \langle e^{-i\pi \lbrack \pm \hat{\cal N}_{A-}(d,\tau )+\hat{\cal N}
_{A+}(d,\tau )]}\rangle \,,  \label{G}
\end{equation}%
where upper sign is taken in the first term of the exponent for $s{=}\uparrow$ and the lower sign for
$s{=}\downarrow$. This expectation value is taken in the stationary scattering state of $\hat{H}$
specified by the temperature and chemical potentials of incident channels,
and the operators 
\begin{equation}
\hat{\cal N}_{\alpha }(d,\tau )=\int_{d}^{d+v_{\pm }\tau }\hat{\Psi}_{\alpha}^{\dagger }(y)\hat{\Psi}_{\alpha}(y)dy  \label{N}
\end{equation}%
count electrons within the spatial intervals ${d}\leq y\leq {
d+v_{\pm }\tau }$ on the channels $A_{\pm }$. 

Equations~(\ref{H}) - (\ref{N}) form the central results of this paper: they
express the observable of interest in an interacting, non-equilibrium
system, in terms of the expectation value of a single-particle operator,
evaluated as an average in a stationary scattering state of a
single-particle Hamiltonian. We now use these equations to discuss the physics of
relaxation in this system and to calculate the tunnelling conductance.

A simple physical picture of the relaxation process, and an identification
of relevant length scales, follows from the form of Eq.~(\ref{G}). This
picture has the same content as our discussion (above and in Ref.~\onlinecite
{quench2009}) of relaxation arising from two plasmon velocities, but is phrased
in terms of the transformed fermions. Note first that the operators $\hat{N}%
_{\pm }(d,\tau )$ count fermions that pass through the QPC in two time
windows, both of duration $\tau $, but with a relative delay of $d/v_{%
\mathrm{eff}}$, where 
\begin{equation}
1/v_{\mathrm{eff}}=1/v^{-}-1/v^{+}=2g/(v^{2}+g^{2}).
\end{equation}%
At $d=0$ the two windows exactly overlap. Then, for example, $G_\uparrow(0,\tau )$
acquires the same contribution from each fermion inside the window,
regardless of tunneling. In consequence, and as expected, the electron distribution at $d=0$
in the $\uparrow $ channel is unaffected by tunneling, since the 
QPC couples $\downarrow$ channels. (Conversely, it is clear that $G_\downarrow(d,\tau )$ is
affected by tunnelling even for $d\to0$, since it depends on the difference and not the sum of the
fermion numbers in the two windows). Far downstream from the
QPC, by contrast, $d/v_{\mathrm{eff}}\gg \tau $, the two windows are widely
separated in time, and the fermion numbers in each are uncorrelated. In this
case, the right side of Eq.~(\ref{G}) is the product of independent factors,
each with the form of a FCS generating function \cite{klich,abanin-levitov}. 
At large  $d$ the tunnelling conductance is the same in both channels. It matches exactly the result
obtained previously \cite{quench2009} by considering a quantum quench, with
the important consequence that even the limiting distribution far from the
QPC is non-thermal, although deviations from a Fermi-Dirac form are small. 
\cite{quench2009} The velocity $v_{\mathrm{eff}}$ can be combined with
voltage $V$ or temperature $T$ to define two lengths: 
\begin{equation}
l_{\mathrm{V}}=\hbar v_{\mathrm{eff}}/e|V|\quad \mathrm{and}\quad l_{\mathrm{T}%
}=\hbar v_{\mathrm{eff}}/2\pi k_{\mathrm{B}}T.
\end{equation}%
These set the scale for relaxation, and diverge in the non-interacting
limit. 

Detailed comparison of our results with the measurements of Ref.~\onlinecite%
{pierre2010} requires an evaluation of the tunnelling conductance at general $d$. For this we
compute $G_s(d,\tau )$ from Eq.~(\ref{G}) 
as summarised in \cite{supplementary}.
The outcome is shown for the two alternative geometries in Figs.~\ref{fig1} and \ref{fig2}.
The theoretical curves match the data well, except for the smallest value of 
$d$ in Fig.~\ref{fig1}. Four parameters enter the calculations: $v_{\mathrm{eff}}$, $T
$, $V$ and the tunneling probability $p$ at the QPC. In addition, one
further parameter is needed to set the energy scale for the data: the gate
voltage-to-energy lever arm $\eta _{\mathrm{G}}$ (defined in Ref.~\onlinecite%
{pierre2010}, supplemental material). Of these parameters, only $v_{\mathrm{%
eff}}$ is unconstrained by independent measurements. We arrive at the fits
in Figs.~\ref{fig1} and \ref{fig2} as follows. We fix $\eta _{\mathrm{G}}=0.062$ and $p\simeq
0.5$ (consistent with the values determined in Refs.~\onlinecite
{pierre2010} and \onlinecite{pierre2009}) and obtain the bath temperature $T=44$ \textrm{%
mK} (close to the experimental value of $T=40$ \textrm{mK}) by fitting data
at zero bias voltage to a thermal Fermi-Dirac distribution. We then compare
numerical results at large $d$ with data measured in the $\downarrow $
channel at the maximum distance from the QPC ($30$ $\mathrm{\mu m}$). We
obtain a good fit at $\beta eV=8$, and use this value for computations shown
in Fig.~\ref{fig1} at all other $d$. Finally, we fix the interaction scale $v_{%
\mathrm{eff}}$ by fitting data at one intermediate distance ($4$ $\mathrm{%
\mu m}$). The theoretical predictions at other distances are then fully
determined. (Our specific value for $p$ [$0.545$ rather than $0.500$] is
significant for the fit only at intermediate $d$ and small energy.) Data for
the geometry of Fig.~\ref{fig2} were taken using a different bias voltage from that
in Fig.~\ref{fig1} and we calculate the theoretical curves shown in Fig.~\ref{fig2} using the
same values of $\eta _{\mathrm{G}}$, $T$ and $v_{\mathrm{eff}}$ as for
Fig.~\ref{fig1}, but scaling the value of $\beta eV$ by the ratio of experimental
voltages. Both values of $\beta eV$ are 25\% lower than the experimental
ones, and this discrepancy is consistent with the `missing energy' reported
in Ref.~\onlinecite{pierre2010}. Possible interpretations of this discrepancy include
loss of energy at the QPC or the existence of additional, neutral edge modes,\cite{pierre2010,buttiker2010,degiovanni2010}  
or invoke \cite{quench2009,LS2011}
the distinction in an interacting system between the measured spectral function and the electron distribution in energy.
The fact that we obtain good fits to all data
except that in Fig.~\ref{fig1} for the smallest $d$ suggests that, if energy is lost
from the edge modes in the experimental system, this probably happens very
close to the QPC. The most significant outcome from this
fitting process is a value for the interaction parameter: $v_{\mathrm{eff}}=6.5\times 10^{4}$ms$^{-1}$. 
A roughly comparable result ($v_{\mathrm{eff}}=10^{5}$ms$^{-1}$)
was obtained from the same data in Ref.\onlinecite{degiovanni2010}, but we believe our
fitting procedure is more precise. Both determinations are in
the range measured for edge state velocities in gated samples.\cite{pierre2010}

An additional quantity of interest is the energy density $\varepsilon ^{s}(d)
$ in channel $s$: it characterises with a single value the electron
distribution \cite{pierre2010}. Moreover, we can evaluate it explicitly from
Eq.~(\ref{G}) because (see Ref.~\onlinecite{quench2009}) it is proportional to $%
\partial _{\tau }G^{\pm }(d,\tau )|_{\tau =0}$ and $G^{\pm }(d,\tau )$ has
an expansion for small $\tau $ in cumulants of $\hat{N}_{\pm }(d,\tau )$.
Interactions cause energy exchange between channels, so that $\varepsilon
^{s}(d)$ evolves with $d$. We obtain 
\begin{equation}
\varepsilon ^{\pm }(d)=\varepsilon _{\mathrm{T}}+\frac{\varepsilon _{\mathrm{%
V}}}{2}\left[ 1\mp \frac{l_{\mathrm{V}}^{2}}{l_{\mathrm{T}}^{2}}\frac{\sin
^{2}d/2l_{\mathrm{V}}}{\sinh ^{2}d/2l_{\mathrm{T}}}\right] \,.
\label{energy}
\end{equation}%
Here $\varepsilon _{\mathrm{V}}\equiv p(1-p){eV}/{4\pi \hbar v}$ is the
excess energy in a single channel due to a double-step electron distribution
generated by a QPC with tunneling probability $p$. Eq.~(\ref{energy}), derived using a different approach in 
Ref.~\onlinecite{degiovanni2010}, shows
that the energy densities interpolate between values just after the QPC of $%
\varepsilon ^{-}(0)=\varepsilon _{\mathrm{T}}+\varepsilon _{\mathrm{V}}$ and 
$\varepsilon ^{+}(0)=\varepsilon _{\mathrm{T}}$, and values at large $d$ of $%
\varepsilon ^{\pm }(d)=\varepsilon _{\mathrm{T}}+\varepsilon _{\mathrm{V}}/2$%
. The relaxation length is set by the smaller of $l_{\mathrm{T}}$ and $l_{%
\mathrm{V}}$. This relaxation is oscillatory, but oscillations are strongly
suppressed if $eV\ll k_{\mathrm{B}}T$. 

Details of calculations are given in Supplementary Material \cite{supplementary}.
We thank F.~Pierre for extensive discussions and for providing access to the
data of Ref.~\onlinecite{pierre2010}. The work was supported in part by EPSRC
grant EP/I032487/1.

\begin{widetext}

\section{Supplementary Material}

\end{widetext}

\section{Model}
\label{model}

We take a Hamiltonian with kinetic, interaction and tunnelling terms, in the
form $\hat{H}=\hat{H}_{\mathrm{kin}}+\hat{H}_{\mathrm{int}}+\hat{H}_{\mathrm{%
tun}}$. Using the labels $\eta =1,2$ to distinguish edges according to their
source, and $s=\uparrow ,\downarrow $ to differentiate between the two
channels on a given edge, the fermion creation operator at point $x$ for
channel $\eta ,s$ is $\hat{\psi}_{\eta s}^{\dagger }(x)$. The density operator is 
$\hat{\rho}_{\eta s}(x)=\hat{\psi}_{\eta s}^{\dagger }(x)\hat{\psi}_{\eta
s}(x)$.
Taking all four channels $\eta ,s$ to have the same bare velocity $v$, and
assuming a contact interaction of strength $g$ between electrons in
different channels on the same edge, we have \cite{Wen_sup} 
\begin{equation}
\hat{H}_{\mathrm{kin}}=-i\hbar v\sum_{\eta ,s}\int \hat{\psi}_{\eta
s}^{\dagger }(x)\partial _{x}\hat{\psi}_{\eta s}(x)\ \mathrm{d}x
\end{equation}%
and 
\begin{equation}
\hat{H}_{\mathrm{int}}=2\pi \hbar g\sum_{\eta }\int \hat{\rho}_{\eta
\uparrow }(x)\hat{\rho}_{\eta \downarrow }(x)\ \mathrm{d}x\,.
\end{equation}%
Tunneling with amplitude $t_{\mathrm{QPC}}$ at the QPC between
channels $1{\downarrow }$ and $2{\downarrow }$ is described by 
\begin{equation}
\hat{H}_{\mathrm{tun}}=t_{\mathrm{QPC}}\hat{\psi}_{1\downarrow }^{\dagger
}(0)\hat{\psi}_{2\downarrow }(0)+\mathrm{h.c.}\,.
\end{equation}%
A bias voltage $V$ generates a chemical potential difference $eV$ between
incident electrons on edge 1 and those on edge 2.

The observable of interest is the tunnelling conductance as a function of energy $E$ and
distance $d>0$ from the QPC, in channel $2,s$. This is the Fourier
transform of the correlator 
\begin{equation}\label{defG_sup}
G_{s}(d,\tau )=\langle e^{i\hat{H}\tau/\hbar} \hat{\psi}_{2s}^{\dagger }(d)e^{-i\hat{H}\tau/\hbar}\hat{\psi}%
_{2s}(d)\rangle ,
\end{equation}%
where the average is taken in the non-equilibrium steady state. The tunnelling conductance is
determined by a measurement of the current $I_{\mathrm{QD}}(E)$ through a
quantum dot with a single level at energy $E$ weakly coupled to the channel.
With tunnelling amplitude $t_{\mathrm{D}}$ to the dot, this is \cite%
{quench2009_sup} 
\begin{equation}
I_{QD}(E)=\frac{e|t_{\mathrm{D}}|^{2}}{\hbar ^{2}}\int G_{s }(d,\tau
)e^{-iE\tau /\hbar }\ d\tau .  \label{IQD_sup}
\end{equation}%

\section{Refermionization}
\label{refermionization}

The steps required to derive Eqns.~(6) - (8) of the main text are as follows.
First we bosonize the Hamiltonian in the standard way \cite{vonDelft},
introducing bosonic fields $\hat{\phi}_{\eta s}(x)$ with commutation
relations 
$
\lbrack \hat{\phi}_{\eta s}(x),\partial _{y}\hat{\phi}_{\eta s^{\prime
}}(x^{\prime })]=-2\pi i\delta _{\eta \eta ^{\prime }}\delta _{ss^{\prime
}}\delta (x-x^{\prime }),
$
and Klein factors $\hat{F}_{\eta s}$, and representing the fermion operators
via the relation 
\begin{equation}\label{bosonize}
\hat{\psi}_{\eta s}(x)= (2\pi a)^{-1/2}\hat{F}_{\eta s}e^{i\frac{2\pi}{L}\hat{N}_{\eta s}}e^{-i\hat{\phi}_{\eta s}(x)},
\end{equation}
where, as usual, $a$ is a short-distance cut-off, $\hat{N}_{\eta s}$ is the fermion number operator, and $L$ is the length of the edge.
After bosonization, the combination $\hat{H}_{\mathrm{kin}}+\hat{H}_{\mathrm{%
int}}$ is diagonalized by a rotation to new bosonic fields $\hat{\chi}%
_{\alpha }$, where $(\hat{\chi}_{S_{+}}\ \hat{\chi}_{A_{-}}\ \hat{\chi}%
_{A_{+}}\ \hat{\chi}_{S_{-}})^{T}=U(\hat{\phi}_{1\uparrow }\ \hat{\phi}%
_{1\downarrow }\ \hat{\phi}_{2\downarrow }\ \hat{\phi}_{2\uparrow })^{T}$
and 
\begin{equation}
U=\frac{1}{2}\left( 
\begin{array}{rrrr}
1 & 1 & 1 & 1 \\ 
1 & -1 & 1 & -1 \\ 
1 & 1 & -1 & -1 \\ 
1 & -1 & -1 & 1%
\end{array}%
\right) \,.  \label{U}
\end{equation}%
Next we use the fields $\hat{\chi}_{\alpha }$ to define new fermion
operators\cite{Fabrizio_sup,zarand_sup} $\hat{\Psi}_{\alpha }$. Crucially, besides
diagonalising $\hat{H}_{\mathrm{kin}}+\hat{H}_{\mathrm{int}}$, the
transformations ensure that $\hat{H}_{\mathrm{tun}}$ remains a
single-particle operator when expressed in terms of $\hat{\Psi}_{\alpha }$.
Specifically, after bosonization 
\begin{equation}
\hat{H}_{\mathrm{tun}}=t_{\mathrm{QPC}}\hat{F}_{1\downarrow }^{\dagger }\hat{%
F}_{2\downarrow }e^{i[\hat{\phi}_{1\downarrow }(0)-\hat{\phi}_{2\downarrow
}(0)]}+\mathrm{h.c.},
\end{equation}%
while under rotation $\hat{\phi}_{1\downarrow }(0)-\hat{\phi}_{2\downarrow
}(0)=\hat{\chi}_{A_{+}}(0)-\hat{\chi}_{A_{-}}(0)$. Since $\hat{\chi}_{A\pm
}(0)$ appear in this expression with unit coefficients, we can introduce new
Klein factors $\hat{F}_{\alpha }$ and new fermion fields $\Psi _{\alpha
}\sim \hat{F}_{\alpha }e^{-i\hat{\chi}_{\alpha }}$ to obtain the expression
for $\hat{H}$ displayed in Eq.~(6) of the main text. The
required transformation of Klein factors has been given previously in the
context of a two-channel Kondo model:\cite{zarand_sup} since fermion number
operators should transform following Eq.~(\ref{U}), we require 
\begin{eqnarray}\label{klein}
\hat{F}_{S_{-}}^{\dagger }\hat{F}_{A_{-}}^{\dagger } &{=}&\hat{F}_{1\uparrow
}^{\dagger }\hat{F}_{1\downarrow },\quad \hat{F}_{S_{-}}\hat{F}_{A_{-}}^{\dagger
}=\hat{F}_{2\downarrow }^{\dagger }\hat{F}_{2\uparrow }, \nonumber \\ 
\hat{F}_{S_{-}}^{\dagger }\hat{F}_{A_{+}}^{\dagger } &=&\hat{F}_{1\uparrow
}^{\dagger }\hat{F}_{2\downarrow },\quad \hat{F}_{S_{+}}^{\dagger }\hat{F}%
_{A_{-}}^{\dagger }=\hat{F}_{1\uparrow }^{\dagger }\hat{F}_{2\downarrow
}^{\dagger }.
\end{eqnarray}%
This implies that the combination appearing in $\hat{H}_{\mathrm{tun}}$
has the transformation $\hat{F}_{1\downarrow }^{\dagger }\hat{F}_{2\downarrow }=-\hat{F}%
_{A_{+}}^{\dagger }\hat{F}_{A_{-}}$. Note that a unit change
in the occupation number of one of the new fermions results in
changes of one half for the occupation numbers of the original fermions.
Physical states in the new basis must therefore satisfy certain selection rules, to ensure
that fermion occupation numbers in the original basis are integer.
These selection rules are set out in Ref.~\onlinecite{zarand_sup}.
For our purposes the key point is that $\hat{H}_{\mathrm{tun}}$ does not
connect the physical and unphysical sectors, because the new fermion operators appear in it in pairs.

We can also transform in this way the operators that are required to generate an
incident state with a density difference between channels. Since $\hat{F}%
_{1\uparrow }^{\dagger }\hat{F}_{1\downarrow }^{\dagger }=\hat{F}%
_{A_{+}}^{\dagger }\hat{F}_{S_{+}}^{\dagger }$ and $\hat{F}_{2\downarrow
}^{\dagger }\hat{F}_{2\uparrow }^{\dagger }=\hat{F}_{A_{+}}\hat{F}%
_{S_{+}}^{\dagger }$, a bias voltage $V$ between channels $1$ and $2$ is
represented by setting the chemical potential to be $eV$ in channel $A_{+}$
and zero in channels $A_{-}$ and $S_{\pm }$.

The task now is to evaluate the correlation function $G_s(d,t)$. For this it
is convenient to
work in the interaction representation, with $\hat{H}_{\rm tun}$ as the `interaction' and
$\hat{H}_0 \equiv \hat{H}_{\rm kin} + \hat{H}_{\rm int}$, using
a superscript $I$ to indicate operators in this representation:
$A^{I}(t)=e^{i\hat{H}_0t}A e^{-i\hat{H}_0t}$. 
Time evolution of the new bosonic and fermionic fields fields is simple in this picture:
we have 
\begin{eqnarray}
\hat{\chi}^{I}_{A\pm}(x,t) &=& \hat{\chi}_{A\pm}(x-v_\pm t)\,,\nonumber \\
\hat{\Psi}^{I}_{A\pm}(x,t) &=& \hat{\Psi}_{A\pm}(x-v_\pm t)\,, \nonumber
\end{eqnarray}
and similarly for $\hat{\chi}^{I}_{S\pm}(x,t)$ and $\hat{\psi}^{I}_{S\pm}(x,t)$.

The time evolution operator in the interaction representation is
\begin{equation}
\hat{S}^{I}(t)=\mathrm{{exp}}\left[-\frac{i}{\hbar }\int_{-\infty }^{t}\hat{H}^{I}_{\rm tun}(\tau ){\rm d}\tau \right]\,,
\end{equation}
(the usual time-ordering is not required here because $[\hat{H}^{I}_{\rm tun}(t_1),\hat{H}^{I}_{\rm tun}(t_2) ]=0$ for all $t_1$ and $t_2$).
Scattering states are generated by the action of $\hat{S}^{I}(t)$: Eq.~(\ref{defG_sup}) takes the form
$G_s(d,\tau) = \langle Q
\rangle_0
$, 
where
\begin{equation}
Q \equiv [\hat{S}^{I}(\tau)]^\dagger [\hat{\psi}^I_{2s}(d,\tau)]^\dagger \hat{S}^I(\tau) [\hat{S}^I(0)]^\dagger \hat{\psi}^I_{2s}(d,0)\hat{S}^{I}(0) \nonumber
\end{equation}
and $\langle \ldots \rangle_0$ denotes a conventional thermal average, with Hamiltonian $\hat{H}_0$ and chemical 
potentials $\mu_1$ and $\mu_2$ on the two edges, defined by
\begin{equation}
\langle \dots \rangle_0 = Z^{-1} {\rm Tr} \left\{e^{-\beta(\hat{H}_0 - \mu_1\hat{N}_1-\mu_2 \hat{N}_2)} \dots  \right\},
\end{equation}
where $Z= {\rm Tr} \{e^{-\beta(\hat{H}_0 - \mu_1\hat{N}_1-\mu_2 \hat{N}_2)} \}$ and $\hat{N}_{1}\equiv \hat{N}_{1\uparrow} + \hat{N}_{1\downarrow}$,
the number operator for edge 1 (and correspondingly for $\hat{N}_2$). 

We now describe how the quantity $Q$ may be simplified. First, by a straightforward though lengthy calculation one can show that
\begin{equation}
[\hat{H}_{\rm tun}^I(t_1),\hat{\psi}^I_{2s}(d,t_2)] =0 \quad {\rm for} \quad t_1 \notin [t_2-d/v_+,t_2-d/v_-]\,.\nonumber
\end{equation}
Next, we insert $[\hat{S}^I(\infty)]^\dagger \hat{S}^I(\infty)$  between the factors of $\hat{S}^I(\tau)$ and $[\hat{S}^I(0)]^\dagger$
in our expression for $Q$. 
We then commute $\hat{S}^I(\tau)[\hat{S}^I(\infty)]^\dagger $ to the left, and $\hat{S}^I(\infty) [\hat{S}^I(0)]^\dagger$ to the right, to obtain
\begin{equation}\label{Q}
Q = [\hat{S}^{I}(\infty)]^\dagger [\hat{\psi}^I_{2s}(d,\tau)]^\dagger  \hat{\psi}^I_{2s}(d,0)\hat{S}^{I}(\infty)\,.
\end{equation}
Bosonizing using Eq.~(\ref{bosonize}) we have
\begin{multline*}
{[\hat{\psi}^I_{2s}(d,\tau)]^\dagger  \hat{\psi}^I_{2s}(d,0)
= (2\pi a)^{-1} e^{i \hat{\phi}_{2s}^I(d,\tau)}e^{-i\frac{2\pi}{L}\hat{N}_{2s}^I(\tau)}}  \\
\times[\hat{F}_{2s}^I(\tau)]^\dagger\hat{F}^I_{2s}(0) e^{i\frac{2\pi}{L}\hat{N}^I_{2s}(0)} e^{-i\phi^I_{2s}(d,0)}\,.
\end{multline*} 
Rotating to the new fields and using  
\begin{equation}
\hat{\chi}^I_\alpha(d,0)  - \hat{\chi}^I_\alpha(d,\tau) + \frac{2\pi}{L}[\hat{N}^I_\alpha(\tau) - \hat{N}^I_\alpha(0)] = 2\pi \hat{\cal N}_\alpha(d,\tau) \nonumber
\end{equation}
with $\hat{\cal N}_\alpha(d,\tau)$ defined in Eq.~(8) of the main text, and setting $\mu_2=0$ so that $\hat{F}_{2s}^I(\tau) = \hat{F}_{2s}$, we find
\begin{multline*}
[\hat{\psi}^I_{2s}(d,\tau)]^\dagger  \hat{\psi}^I_{2s}(d,0)=\\
(2\pi a)^{-1}
e^{-i\pi [({\cal N}_{A+}(d,\tau) \pm {\cal N}_{A-}(d,\tau)) - ( {\cal N}_{S+}(d,\tau)\pm {\cal N}_{S-}(d,\tau)]}
\end{multline*}
where the signs $\pm$ correspond to the choices $s=\uparrow,\downarrow$.
Since $\hat{S}^{I}(t)$ transforms only the channels $A\pm$ and not $S\pm$, we can extract a normalisation factor to arrive at
\begin{equation}\label{ratio}
G_s(d,\tau) = G_0(\tau) \frac{\langle e^{-i\pi [{\cal N}_{A+}(d,\tau) \pm {\cal N}_{A-}(d,\tau)]}\rangle}{\langle e^{-i\pi [{\cal N}_{A+}(d,\tau) \pm {\cal N}_{A-}(d,\tau)]}\rangle_0}
\end{equation}
where $\langle \ldots \rangle \equiv \langle  [\hat{S}^{I}(\infty)]^\dagger \ldots   \hat{S}^{I}(\infty)  \rangle_0$ and 
\begin{eqnarray}
G_0(\tau) &\equiv& \langle \hat{\psi}_{2s}^\dagger(d,\tau) \hat{\psi}_{2s}(d,0) \rangle_0 \nonumber\\
&=&\frac{i   }{2\beta \hbar (v_{+}v_{-})^{\frac{1}{2}}}\times \frac{1}{\sinh
^{\frac{1}{2}}[\frac{\pi }{\beta \hbar v_{+}}(-v_{+}\tau +ia)]}  \nonumber \\
&&\qquad\times \frac{1}{\sinh ^{\frac{1}{2}}[\frac{\pi }{\beta \hbar v_{-}}
(-v_{-}\tau +ia)]}\,.
\end{eqnarray}

A final step is to consider the effect of the evolution operators $[\hat{S}^{I}(\infty)]^\dagger$ and $\hat{S}^{I}(\infty)$ 
on the basis states in which the expectation value is calculated. As these states are generated by the action of the operators 
$\hat{\Psi}_{A\pm}(x)$ on the vacuum, we must find how these operators transform. We do this
by solving the Schr\"odinger equation with the Hamiltonian of Eq.~(6) of the main text. The solution involves the scattering amplitudes at the QPC:
introducing $\theta =t_{\mathrm{QPC}}/\hbar \sqrt{v_{+}v_{-}}$, the tunnelling and reflection probabilities 
are $\sin ^{2}\theta $ and $\cos^{2}\theta$, while 
\begin{multline}\label{trans1}
[\hat{S}^{I}(\infty)]^\dagger \hat{\Psi}_{A+}(x)\hat{S}^{I}(\infty) =\\
\cos \theta \, \hat{\Psi}_{A+}(x) -i\sin \theta \, \left[v_{-}/v_{+}\right]^{{1}/{2}} 
\hat{\Psi}_{A_{-}}^{\dagger } (v_{-}x/v_{+})
\end{multline}
and
\begin{multline}\label{trans2}
[\hat{S}^{I}(\infty)]^\dagger \hat{\Psi}_{A-}(x)\hat{S}^{I}(\infty) =\\
\cos \theta \, \hat{\Psi}_{A-}(x) -i\sin \theta \, \left[v_{+}/v_{-}\right]^{{1}/{2}} 
\hat{\Psi}_{A_{+}}^{\dagger } (v_{+}x/v_{-})\,.
\end{multline}
Hence the average $\langle \dots \rangle$ in a scattering state is evaluated by combining 
the rotation between channels defined in Eqns.~(\ref{trans1}) and (\ref{trans2}) 
with the thermal average $\langle \ldots \rangle_0$.

\section{Numerical evaluation of the tunnelling conductance}
\label{numerics}

In the following we outline the methods we use for numerical evaluation of the
correlation function defined in Eq.~(7) of the main text, and hence the differential tunnelling conductance
shown in Figs.~1 and 2 of the main text. The central problem is to evaluate the normalised
expectation value appearing on the right hand side of Eq.~(\ref{ratio}), which has the form
\begin{equation}
\langle X \rangle_{\rm norm} \equiv {\langle X \rangle}/{\langle X \rangle_0}
\end{equation}
with $X=e^{-i\pi [{\cal N}_{A+}(d,\tau) \pm {\cal N}_{A-}(d,\tau)]}$.
In the limit of large distance $d$ from the QPC,
contributions to the two time windows appearing in the exponent in $X$ are independent and
it is convenient to use the approach described in Ref.~\onlinecite{mirlin_sup}. At finite $d$
there is no such factorisation and we resort instead to a method
similar to one described in Ref.\onlinecite{quench2009_sup},
appropriately modified to take account of finite temperature.

(i) \textit{Long distance limit.} At large $d$ the expectation value
factorizes into a product of two functions, each of which count number of
particles in a fixed time window, so that
\begin{equation}
\langle e^{-i\pi [{\cal N}_{A+}(d,\tau) \pm {\cal N}_{A-}(d,\tau)]} \rangle_{\rm norm} = \chi _{- }(\mp \pi ,\tau )\chi _{+}(-\pi ,\tau )\nonumber
\end{equation}
with
\begin{equation}
\chi _{\pm }(\delta ,\tau )=\langle e^{i\delta \hat{\cal N}_{\pm }(d,\tau )}\rangle_{\rm norm}
.  \label{FCS}
\end{equation}
The functions $\chi _{\pm }(\delta ,\tau )$ have a FCS form and can be written in terms of
determinants for non-interacting fermions. One has\cite{klich_sup}
\begin{equation}
\langle e^{i\delta \hat{\cal N}_{\pm }(d,\tau )}\rangle
=\det \{[1-\hat{P}(e^{-i\delta }-1)\hat{n}_{\pm
}(\varepsilon )\hat{P}]\}\,.  \label{det2}
\end{equation}%
Here $n_{\pm }(\varepsilon )$ is the corresponding electron energy
distribution in a given channel at finite temperature and bias voltage after the action of $\hat{S}^I({\infty})$ 
(i.e. a double-step) while $\hat{P}$ is a projection operator that is diagonal in the time
domain, having the action on a time-dependent
function $y(t)$: $\hat{P}y(t)=y(t)$ if $t\in \lbrack 0,\tau ]$ and $\hat{P}%
y(t)=0$ otherwise. Using the regularization procedure proposed in Ref. \onlinecite
{mirlin_sup} we can write the determinant (\ref{det2}) in the form%
\begin{equation}
\langle e^{i\delta \hat{N}_{\pm }(\tau )}\rangle
=\det [f(t_{i}-t_{j})]\,,  \label{chi}
\end{equation}%
where the function $f(t)$ is defined as a Fourier transform of 
\begin{equation}
\tilde{f}(\varepsilon )=[1-n_\pm(\varepsilon )(e^{-i\delta }-1)]e^{-i\frac{\delta }{2}
\frac{\varepsilon }{\Lambda }},
\end{equation}
which is periodic in the domain $[-\Lambda ,\Lambda ]$, with $\Lambda $
a high energy cutoff on single-particle states and $t_{j}=j\pi /\Lambda .$
Note that the phase factor $e^{-i\frac{\delta }{2}\frac{\varepsilon }{\Lambda }}$ appearing here is crucial, since without it
one would wrongly obtain a result periodic in $\delta$.

\begin{figure}[tbp]
\epsfig{file=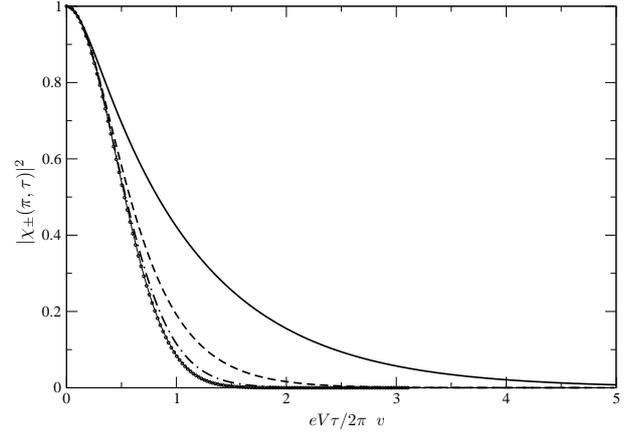,width=8cm}
\caption{The quantity $|\chi _{\pm }(\pi ,\tau )|^2$ [see Eq.~(\protect\ref{FCS})] as a
function of time $\tau$ for different values of parameter $\protect\beta eV$ at
tunneling probability $p=0.5$ Thick solid line: $\protect\beta eV=2$; dashed line: $\protect\beta eV=5$; 
dot-dashed line: $\protect\beta eV=10$; circles: $\protect
\beta eV=100$. Thin solid line corresponds to asymptotic behaviour at large $
\protect\beta eV$, which is represented within numerical errors by the function $\exp [-\frac{1}{12}\protect\pi ^{2}\protect\tau 
^{2}(k_{B}T_{\mathrm{ds}})^2/\hbar ^{2}].$}
\label{m3}
\end{figure}

\begin{figure}[tbp]
\epsfig{file=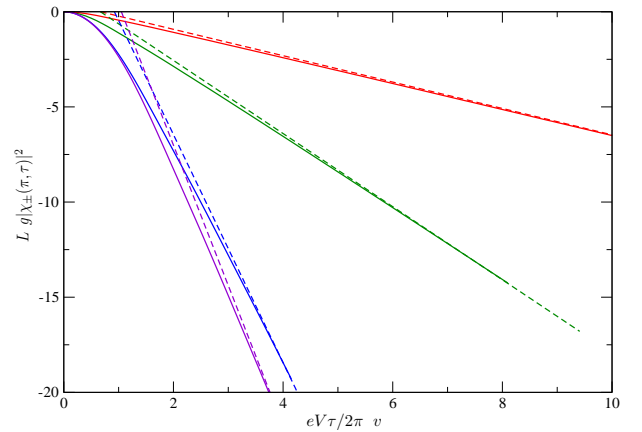,width=8cm}
\caption{(Color online). Comparison of the asymptotic behaviour of
$|\chi _{\pm}(\pi ,\tau )|^2$ [see Eq.~(\protect\ref{FCS})] from Ref.\protect\onlinecite{mirlin_sup},
(solid lines) with results obtained numerically using Eq.~(\protect\ref%
{chi}) (dashed lines), at tunneling probability $p=0.5$ for different values of $\delta$. 
From top right to bottom left: red lines $%
\protect\delta =\protect\pi /2,$ green lines ($\protect\delta =3\protect\pi /4$),
blue lines ($\protect\delta =\protect\pi -0.1$) and violet lines ($\protect\delta =%
\protect\pi -0.01$).}
\label{mirlin}
\end{figure}

The normalisation $\langle e^{i\delta \hat{\cal N}_{\pm }(d,\tau )}\rangle_0$ is obtained from similar
expressions in which the non-equilibrium distribution $n_\pm(\varepsilon)$ is replaced by a thermal one, and
in practice we evaluate directly the ratio $\chi _{\pm }(\delta ,\tau)$.
We check numerical convergence by changing the value of the cutoff $\Lambda .$ The size of the
matrices required grows linearly with $\tau$, the largest necessary being $%
2000\times 2000.$ Some results obtained using this approach are shown in Figs.~\ref{m3} and \ref{mirlin}.
The numerical calculations can be tested at large $\tau$ by comparison with the asymptotic analytic results derived in
Ref.~\onlinecite{mirlin_sup}: as shown in Fig.~\ref{m3} the agreement is excellent. An incidental by-product of our 
calculations is the discovery of an interesting new feature of the asymptotic behaviour. According to Ref.~\onlinecite{mirlin_sup}
this is exponential in $\tau$, with a rate that diverges for $\delta=\pi$ and $p=1/2$. In fact we find numerically
that decay at these parameter values is not exponential but Gaussian in $\tau$. It would be interesting to search for an
analytical derivation of that form. 

(ii) \textit{Finite distance. }At finite distance $d$ the correlation function
$G_s(d,\tau)$ is not simply a product of independent contributions from the two channels 
$A\pm$, and so cannot be expressed in terms of quantities familiar from the theory of FCS.
We require instead a different numerical approach. Using Levitov's determinant formula \cite{klich_sup}, we have
\begin{multline}
\langle  [\hat{S}^{I}(\infty)]^\dagger 
e^{-i\pi [{\cal N}_{A+}(d,\tau) \pm {\cal N}_{A-}(d,\tau)]}
   \hat{S}^{I}(\infty)  \rangle_0 \nonumber\\
= \det [1-n(\varepsilon)[[\hat{S}(\infty)]^{\dagger }e^{-i\pi \lbrack \pm 
\hat{\cal N}_{A-}(d,\tau )+\hat{\cal N}_{A+}(d,\tau )]}\hat{S}^I(\infty)-1].  \label{GF_k}
\end{multline}
Here the determinant is in the two-channel Fock space and $n(\varepsilon)$
is a Fermi-Dirac distribution in each channel, but with two distinct chemical potentials.
To evaluate this determinant we express $\hat{\cal N}_{A\pm}(d,\tau)$ as bilinears in the
fermion creation and annihilation operators $\hat{\Psi}^{\dagger}_{A\pm}(x)$ and  $\hat{\Psi}_{A \pm}(x)$.
We evaluate this determinant in a basis of eigenstates of $\hat{H}_{\rm kin}$, considering 
edges of finite length with periodic boundary conditions and imposing energy cut-offs to
obtain a matrix of finite size. For adequate convergence with increasing system size, we find that
it is necessary to scale the lengths $L_\pm$ of the two channels according to the velocity, setting
$v_+/L_+ = v_-/L_-$. With this choice, a basis of a few thousand states is sufficient to obtain the results
we present here. 

\begin{figure}[tbp]
\epsfig{file=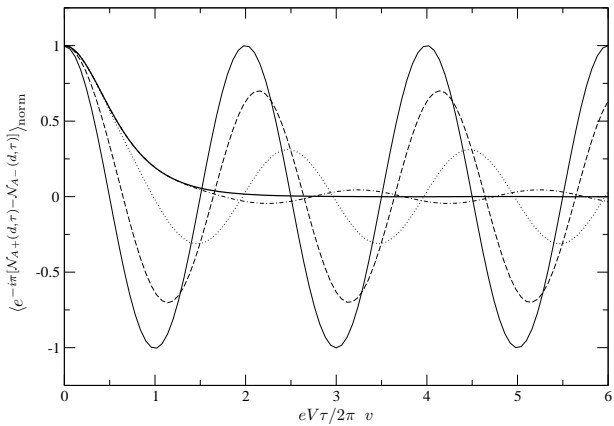,width=8cm}
\caption{Time dependence of the normalised correlator $\langle e^{-i\pi [{\cal N}_{A+}(d,\tau) - {\cal N}_{A-}(d,\tau)]} \rangle_{\rm norm}$ 
for the \textit{same} channel as that coupled by the QPC. Curves are for tunnelling probability $p=1/2$, $eV/k_{\rm B}T = 5$ and different 
distances $d$ from the QPC: (thin solid line) $d/l_{v}=0$; (dashed line) $d/l_{v}=2.5$; 
(dotted line) $d/l_{v}=5$; (dot-dashed line) $d/l_{v}=10$; and (thick solid line) $x/l_{v}=20$.\label{same}}
\end{figure}

\begin{figure}[b]
\epsfig{file=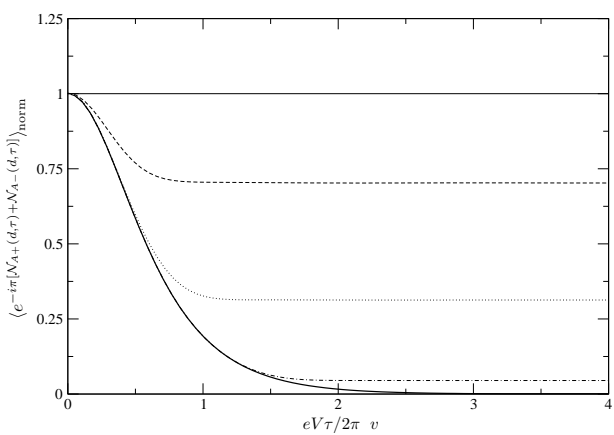,width=8cm}
\caption{Time dependence of the normalised correlator $\langle e^{-i\pi [{\cal N}_{A+}(d,\tau) + {\cal N}_{A-}(d,\tau)]} \rangle_{\rm norm} $ 
for the channel \textit{not} coupled by the QPC. Other parameters as for Fig.~\ref{same}.\label{opposite}}
\end{figure}

We show representative results for the normalised correlator $\langle e^{-i\pi [{\cal N}_{A+}(d,\tau) \pm {\cal N}_{A-}(d,\tau)]} \rangle_{\rm norm}$
at finite $d$ in Figs.~\ref{same} and \ref{opposite}. The most distinctive feature is an oscillatory time dependence
of the correlator in the channel that is coupled by the QPC, which at $d=0$ and $T=0$ can be obtained exactly as
$\langle e^{-i\pi [{\cal N}_{A+}(0,\tau) \pm {\cal N}_{A-}(0,\tau)]} \rangle_{\rm norm}=\cos (\frac{1}{2}eV\tau/2\protect\pi \hbar )$.

\end{document}